\documentclass[10pt]{iopart}

\usepackage{epsfig}
\usepackage{graphicx}

\begin{document}

\title{Electronic states and magnetic excitations in $\rm{Li V_2 O_4}$:
Exact diagonalization study.}

\author{ S. Burdin$^{a}$, 
N.B. Perkins$^{b,c}$\footnote[3]{Corresponding author: perkins@lnf.infn.it}, 
C. Lacroix$^{d}$
}
\address{
$^a$ Institut Laue-Langevin, B.P. 156, 38042 Grenoble Cedex 9, France \\
$^b$ Joint Institute for Nuclear Research, Dubna, Moscow region, 
141980, Russia \\
$^c$ INFN Laboratori Nazionali di Frascati, c.p. 13, I-00044 Frascati,
Italy \\
$^d$ Laboratoire Louis Neel, CNRS, BP 166, 38042, Grenoble Cedex 9,
France }
\begin{abstract}
Motivated by recent inelastic neutron scattering experiment we examine 
magnetic properties of $\rm{Li V_2 O_4}$. We consider a model which describes 
the  half-filled localized $A_{1g}$ spins interacting   via frustrated 
antiferromagnetic Heisenberg exchange and coupled by local Hund's interaction 
with the $1/8$-filled itinerant $E_g$ band, and study it within an exact 
diagonalization scheme. In the present study we limited the analysis to the 
case of the cluster of two isolated tetrahedrons. We obtained that both the 
ground state structure and low-lying excitations depend strongly on the value 
of the  Hund's coupling which favors the triplet states. With increasing 
temperature the triplet states  become more and more populated which results 
in the formation of non-zero residual magnetic moment. We present the 
temperature dependence of calculated magnetic moment and of the spin-spin 
correlation functions at different values of  Hund's coupling and compare them 
with  the experimental  results.
\end{abstract}
\section{Introduction}
In recent years, there has been a great interest in the study of the origin of 
the heavy fermion (HF) behavior observed in the paramagnetic transition-metal 
oxide  $\rm{Li V_2 O_4}$. This compound has a spinel structure and its 
pyrochlore lattice consists of corner-shared tetrahedrons of ${\rm V}^{+3.5}$ 
ions located in slightly distorted oxygen octahedron. It is the first metal 
showing heavy fermion behavior without any $f$ orbital~\cite{kondo}. The 
electronic specific heat coefficient $\gamma=C_e/T=0.42$ J/Mol K$^2$ at 
$T = 1.0$ K is  the highest value measured for 3d transition-metal oxides.

In addition to having quasi-particles with an unusually heavy effective mass, 
$\rm{Li V_2 O_4}$ has some peculiar magnetic properties. The magnetic 
susceptibility and inelastic neutron scattering measurements indicate a spin 
liquid behavior over a large range of intermediate 
temperatures~\cite{ueda,broholm}. The absence of magnetic order indicates that 
geometrical frustration, due to the pyrochore structure, is relevant in this 
system~\cite{theory}. On the other hand, the recent  neutron scattering 
experiments~\cite{murani} reveal that in addition to antiferromagnetic 
correlations, ferromagnetic-like correlations on V sites appear over some  
temperature range. These observations are inconsistent with the 'classical' 
spin-liquid picture, and demonstrate that itinerant contribution of strongly 
correlated electrons, which could lead to the effective ferromagnetic exchange 
interaction between localized spins is also important.

There were some suggestions to explain HF behavior in $\rm{Li V_2  O_4}$ using 
analogy with $4f-$systems where these effects are attributed to the 
hybridization of localized  $4f-$levels and itinerant $spd$ bands. This 
analogy has been based on  the fact that according to the band structure 
calculations~\cite{eyert,anisimov} the trigonal distortion splits $t_{2g}$ 
orbitals into singlet  $A_{1g}$ and doublet $E_{g}$ states, and the center of 
$A_{1g}$ band  lies lower than that of $E_{g}$ band. Therefore, one can make 
a  plausible assumption and treat $A_{1g}$ level as being occupied by a 
localized electron, whereas consider  $E_{g}$ doublet as a quarter filled 
conducting band~\cite{anisimov}. 

In this paper we study the magnetic properties of $\rm{Li V_2 O_4}$ within 
the exact diagonalization analysis of a small cluster. The model that we 
consider includes both purely Heisenberg-like contribution from the 
super-exchange interaction among localized spins and an effective 
ferromagnetic double-exchange contribution driven by the itinerant electronic 
excitations. \\
In order to capture some charge fluctuation of the system, we solve the model 
exactly for the cluster consisting of two disconnected tetrahedrons with 
either 5 and 7 electrons, or only with 6 electrons, projecting out higher 
energy states with 4 and 8 electrons.  
\section{The Model}
In order to consider some charge fluctuations, we study  a cluster of two 
isolated tetrahedrons, i.e. we don't take into account either the 
super-exchange  interaction between localized spins of different tetrahedrons 
or the hopping of electrons between them. For this cluster, we can write the 
total Hamiltonian as $H=H_1 + H_2$, where $H_{1(2)}$ is defined on a single 
tetrahedron: 
\begin{equation}
H_a=\sum_{ij,\alpha\beta\sigma}
t_{ij}^{\alpha\beta}
\left[c_{i\sigma\alpha }^{a\dagger}c_{j\sigma\beta}^{a}+H.c.\right]
-J_H\sum_{i}
{\bf S}_{i}^{a}{\bf \sigma}_{i}^{a}+
J\sum_{i\neq j}{\bf S}_{i}^{a}{\bf S}_{j}^{a}. 
\label{hamiltonian}
\end{equation}
Here $a=1,2$ is the tetrahedron's index; ${\bf S}_{i}^{a}$ are 
$\frac{1}{2}$-spins representing the localized $A_{1g}$ electrons; 
$c_{i\sigma\alpha}^{a\dagger}$ ($c_{j\sigma\beta}^{a}$) the creation 
(annihilation) operator of an itinerant electron with spin 
$\sigma=\uparrow,\downarrow$ and orbital $\alpha=1,2$, corresponding to the 
$E_g$-doublet. \\
The first term in the Hamiltonian Eq.~(\ref{hamiltonian}) describes the 
electron hopping between the nearest neighbor V ions, $t_{ij}^{\alpha  \beta}$ 
being the transfer amplitude. The second term concerns the Hund's coupling 
$J_H$ between the localized spins ${\bf S}_i$ and the local spin density 
${\bf \sigma}_i$ of the itinerant $E_g$-electrons. Finally, the last term 
describes the nearest-neighbor antiferromagnetic super-exchange interaction 
$J$ among localized spins. Here we consider infinite on-site Coulomb repulsion 
between itinerant electrons and project out states with double occupancy.

Considering only direct overlap between $3d$ wave functions 
[$(dd\sigma )=-0.281$ eV, $(dd\pi )=0.0076$ eV]~\cite{mattheiss}, transfer 
matrix elements $t_{ij}^{\alpha \beta}$ can be easily calculated using the 
Table I of the paper by Slater and Koster~\cite{slater}. \\
In further analysis, we consider the value of Heisenberg exchange coupling
$J=10$ meV which is in agreement with the estimates given in the 
literature~\cite{fulde,laad}. The Hund's coupling $J_H$ is known to be little 
screened in solids, and  can  be simply related to its atomic value. The 
estimates reported in the literature suggests that $J_H=0.68-1.0$ 
eV~\cite{tanabe,mizokawa,anisimov,nekrasov}. We use in our study the value 
$J_H=0.8$ eV obtained by the LDA+U ab initio calculations~\cite{anisimov}, 
however we also consider the variation of $J_H$ over a wider range.
\subsection{Hilbert space sectors}
To define the Hilbert space, we choose a basis in which each state $|n\rangle$ 
is a product 
\begin{eqnarray}
|n \rangle  =|N_2, {\cal S}_2, \alpha_2\rangle_2
\otimes |N_1, {\cal S}_1, \alpha_1\rangle_1 , 
\label{state}
\end{eqnarray}
where the state $|\cdots \rangle_a$ characterizes the tetrahedron $a=1,2$. 
The index $N_a$ is the number of electrons (localized plus itinerant) for the 
tetrahedron $a$, with an averaged charge of 6 electrons per tetrahedron 
$N_{1}+N_{2}=12$. For a given tetrahedron, the spin state 
${\cal S}_a\equiv \sum_{i=1}^{4}{\bf S}_{i}^{a}$ characterizes the total spin 
of the four localized electrons. These four spins can be coupled either to 
singlet, triplet or quintet states. If Hund coupling is neglected, considering 
only the antiferromagnetic super-exchange interaction, the ground state 
consists of two-fold degenerate singlet with energy $-3J$, followed by 
ninefold degenerate triplet states and five quintet states with the energy 
$3J$. The quantum number $\alpha_a$ describes all the remaining degrees of 
freedom, in our case, the spin state and the orbital and site occupancy of the 
itinerant electrons on the considered tetrahedron. 

In order to calculate both the ground state and the finite-temperature 
properties of the cluster of two isolated tetrahedrons, first, we performed a 
numerical exact diagonalization of the Hamiltonian $H_1$ characterizing one 
tetrahedron. Because of large degeneracy, the Hilbert space of one tetrahedron 
consists of $256$, $1536$ and $4096$ states for $5$, $6$ and $7$ electrons, 
respectively. However, since the Hamiltonian is invariant under rotation of 
the total spin, it is also diagonal by blocks in the basis of the eigenstates 
of the total spin (itinerant plus localized). 
\subsection{Probabilities}
For a given temperature $T\equiv 1/\beta$, the average value of an operator 
${\Xi}$ (e.g. the nearest neighbors spin-spin correlation) is defined by the 
usual relation 
\begin{equation}
\langle \Xi\rangle
=\frac{1}{\cal Z}\sum_{n}
\langle n|exp(-\beta H)\Xi|n\rangle .
\end{equation}
Choosing a basis diagonalizing either the operator ${\Xi}$ or the Hamiltonian, 
this relation can be rewritten as 
\begin{equation}
\langle \Xi\rangle
=\sum_{n}
p(n,T)
\langle n|\Xi|n\rangle \cdot
\end{equation}
Here, the density matrix 
$p(n,T)\equiv \frac{1}{\cal Z}\langle n|exp(-\beta H)|n\rangle$ is the 
probability for the state $|n\rangle$ to be occupied at temperature $T$. 
Using the definition Eq.~(\ref{state}) of the state $|n\rangle$ of the cluster 
of two isolated tetrahedrons, the probability $P(N, {\cal S})$ that the 
tetrahedron $1$ occupies a state with $N$ electrons and a local spin state 
${\cal S}$ is: 
\begin{eqnarray}
\hspace{-2.2cm}
P(N,{\cal S})&
\hspace{-1cm}
=&
\frac{1}{\cal Z}
\sum_{\alpha_1, \alpha_2, {\cal S}_2}
\langle N,{\cal S}, \alpha_1 |
\langle 12-N,{\cal S}_2,\alpha_2 |
exp(-\beta H)
| 12-N,{\cal S}_2,\alpha_2 \rangle
| N,{\cal S},\alpha_1 \rangle
\nonumber\\
&
\hspace{-1cm}
=&
\frac{{\cal Z}_{12-N}}{\cal Z}
\sum_{\alpha_1}
\langle N,{\cal S},\alpha_1|
exp(-\beta H_{1})
| N,{\cal S},\alpha_1\rangle,
\end{eqnarray}
where the partition function for {\it one} isolated tetrahedron with $N=5$, 
6 or 7 electrons is defined as 
${\cal Z}_N\equiv\sum_{{\cal S}, \alpha}\langle N,{\cal S},\alpha|
exp(-\beta H_1)| N,{\cal S},\alpha\rangle$.
The partition function of the two tetrahedrons system can be cast into
\begin{equation}
{\cal Z}=
{\cal Z}_5 {\cal Z}_7
+
{\cal Z}_6 {\cal Z}_6
+
{\cal Z}_7 {\cal Z}_5. 
\end{equation}
For the further analysis, it is convenient also to define the probability 
that the tetrahedron $1$ occupies a state with a total local spin ${\cal S}$: 
\begin{equation}
P_{spin}({\cal S})=\sum_{N}
P(N,{\cal S}),
\label{spin}
\end{equation}
and the probability to occupy a state with $N$ electrons: 
\begin{equation}
P_{charge}(N)=\sum_{{\cal S}}P(N,{\cal S}). 
\label{charge}
\end{equation}
\section{Results and discussion.}
\subsection{Low energy states}
\begin{figure}
\vspace{0.5cm}
\includegraphics[angle=0, width=0.53\textwidth]{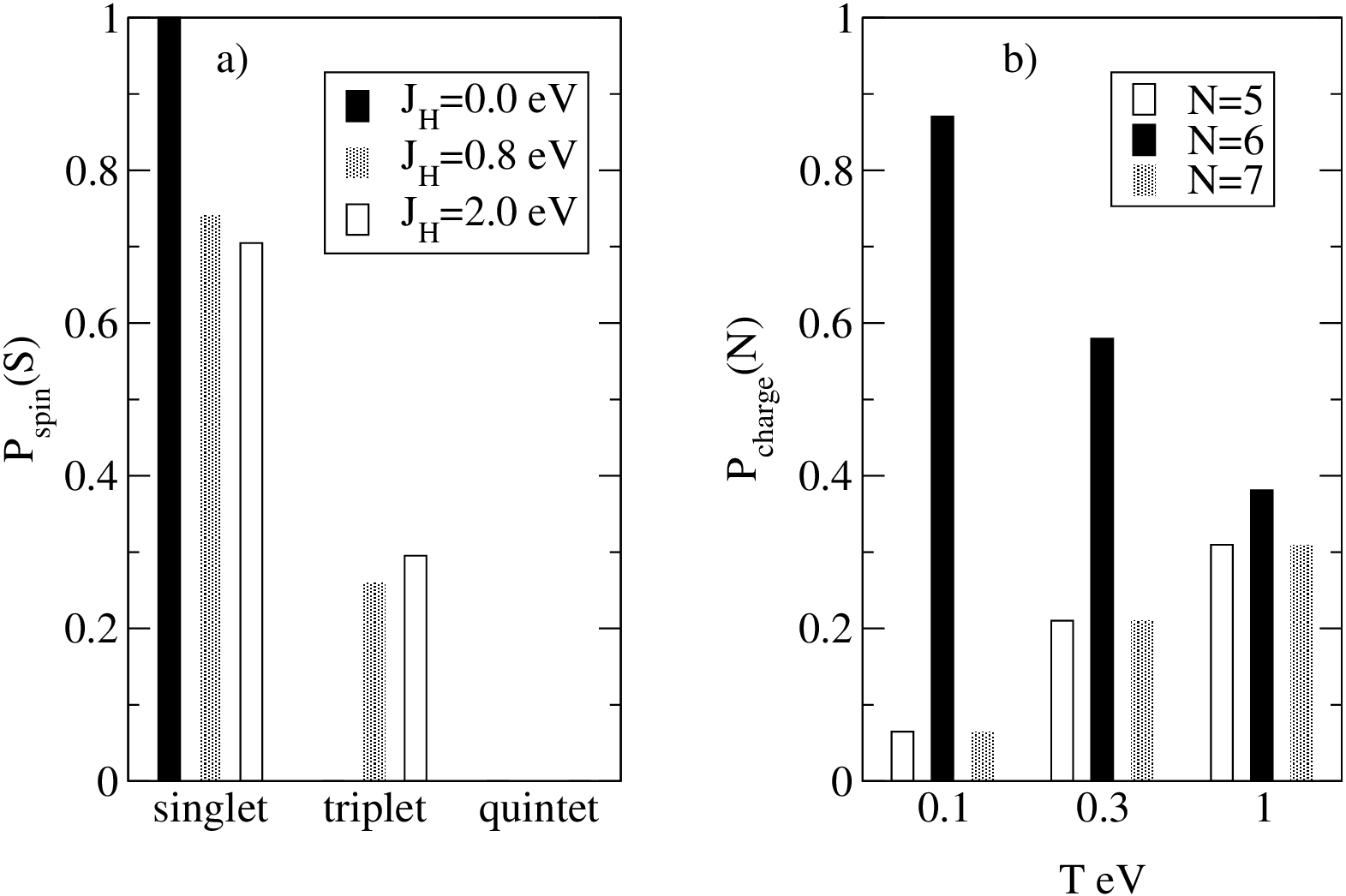}
\hspace{0.05\textwidth}
\includegraphics[angle=0, width=0.47\textwidth]{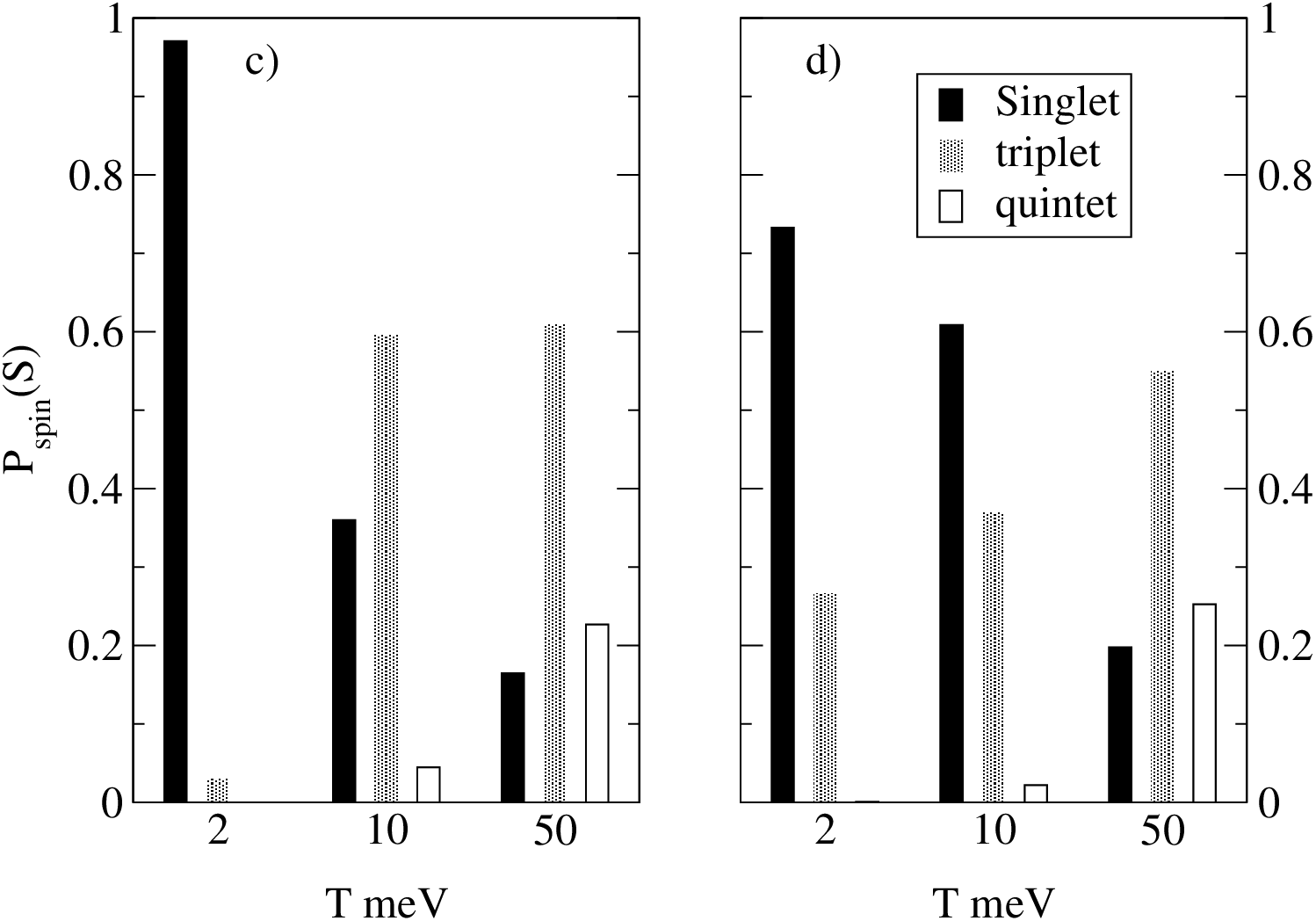}
\caption{\label{proba} 
a) Ground state spin probabilities at different values of Hund's coupling 
$J_H$. 
b) Charge probabilities at temperatures $T=0.1, 0.3, 1.0$ eV for $J_H=0.8$ eV. 
c) and d) Spin probabilities at temperatures $T=2, 10, 50$ meV for $J_H=0.0$ 
eV and $J_H=0.8$ eV, respectively. }
\end{figure}
First, let us discuss the weight of the different spin subsectors in the 
ground state. 
Fig.~\ref{proba}~a) illustrates the spin probabilities 
$P(N=6, {\cal S})$ for $T=0$ and different values of Hund's coupling. At 
$J_H=0.0$ eV (decoupled spins and conducting electrons), the ground state is a 
two-fold degenerate singlet. As soon as the Hund's coupling is switched on, 
the spin degeneracy of the ground state is lifted and some triplet components 
occur. The weight of the triplet states is more than $20\%$ for a realistic 
coupling $J_H=0.8$ eV, but the singlet components remain important even for 
much larger coupling $J_H=2.0$ eV. For the three values of coupling that we 
considered, the quintet contributions to the ground state are negligible. 
In Fig.~\ref{proba}~a) we present only the probabilities of states 
corresponding to the tetrahedron with $N=6$ electrons because the states with 
5 and 7 electrons are higher in energy and, therefore, all the probabilities 
$P(5,{\cal S})$ and $P(7,{\cal S})$ are equal to zero. At low temperatures, as 
in the ground state, mainly the states with $N=6$ electrons are occupied. The 
charge sectors with $N=5$ and $N=7$ contribute only at rather high 
temperatures. In Fig.~\ref{proba}~b) the temperature evolutions of 
$P_{charge}(N)$ are presented for $J_H=0.8$ eV. However, we should note that 
only static charge fluctuations are taken into account here, and considering 
interacting tetrahedrons would certainly increase the charge fluctuations 
contribution at low temperatures.

The evolution of the spin probabilities $P_{spin}({\cal S})$ with increasing 
temperature are presented  in Fig.~\ref{proba}~c) and d) for $J_H=0.0$ eV 
and $J_H=0.8$ eV, respectively. The weights of the triplet states become 
substantial at $T\simeq 10$ meV, and the spin probabilities are mainly 
proportional to the spin degeneracies at $T\simeq 50$ meV. 

\begin{figure}
\vspace{0.5cm}
\includegraphics[angle=0, width=0.5\textwidth]{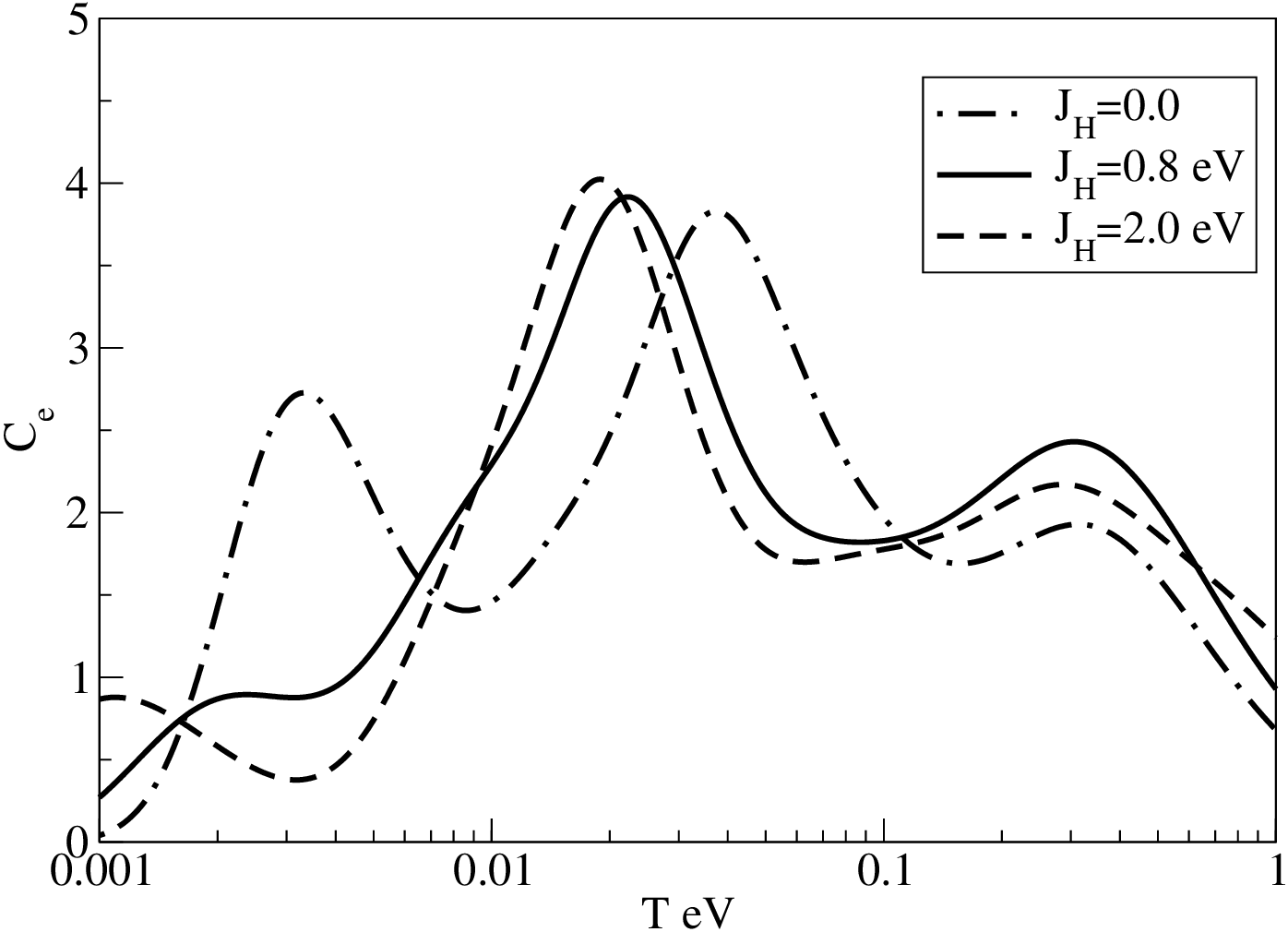}
\hspace{0.05\textwidth}
\includegraphics[angle=0, width=0.5\textwidth]{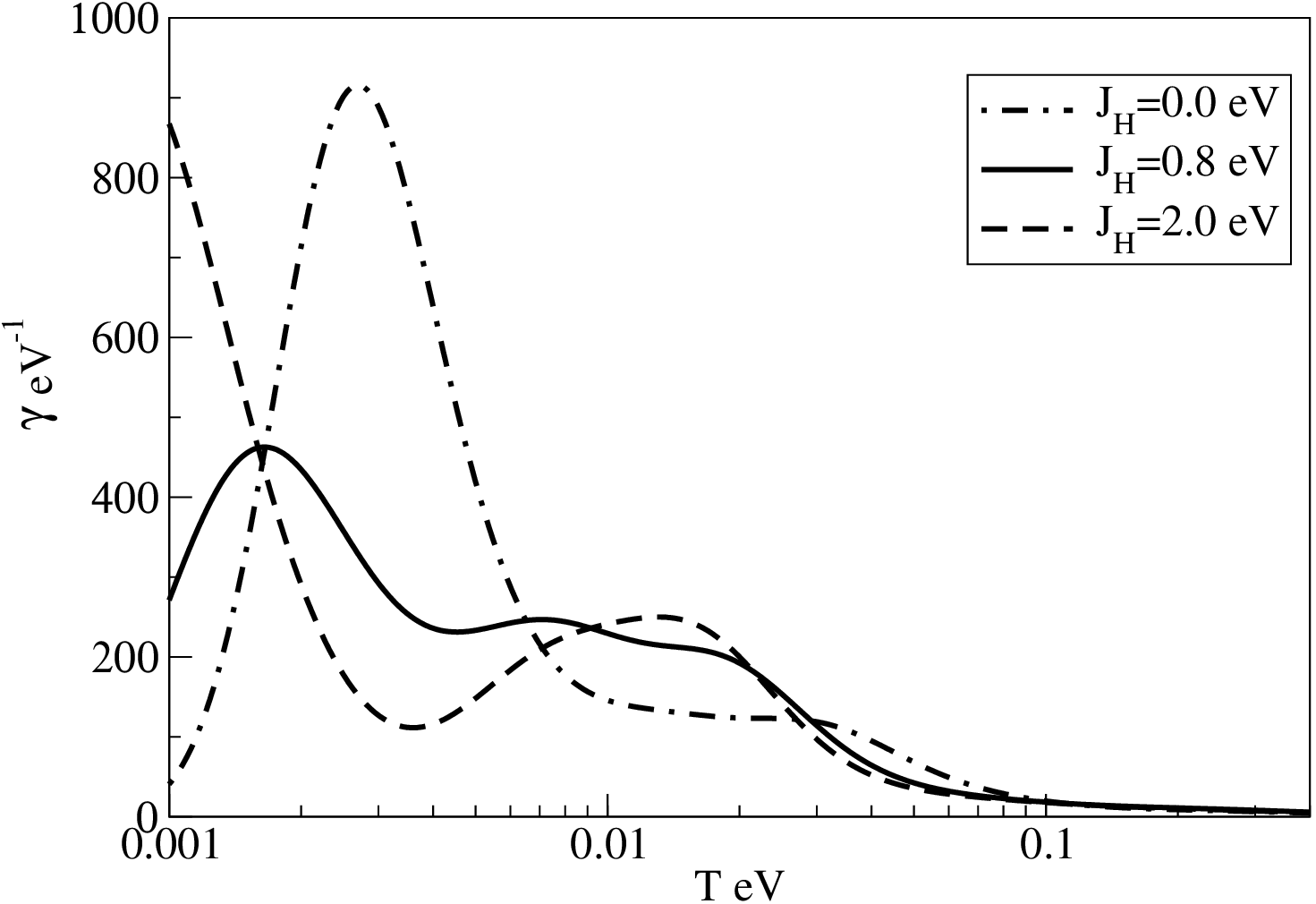}
\caption{\label{thermo}
The temperature dependence of a) the specific heat $C_e$ and b) the 
coefficient  $\gamma=C_e/T$. On both a) and b), the temperature axis is 
logarithmic. }
\end{figure}

To conclude the description of the low energy excitations, let us discuss the 
thermodynamical properties such as the electronic specific heat $C_e$ and the 
coefficient $\gamma=C_e/T$. In Fig.~\ref{thermo}~a) and b) we present the 
temperature dependence of $C_e$ and $\gamma$, respectively. For all values of 
Hund's coupling, the specific heat has three peaks at finite temperatures. Two 
low-energy peaks correspond to spin excitations. For a non zero Hund's 
coupling, the energies of the magnetic states are decreased, and the 
corresponding peak occurs at a lower temperature. The third and the highest 
temperature peak characterizes the charge sector excitations  and it is 
independent of the Hund's coupling. 
\subsection{Local moment and magnetic correlation functions}
\begin{figure}
\vspace{0.5cm}
\includegraphics[angle=0, width=0.5\textwidth]{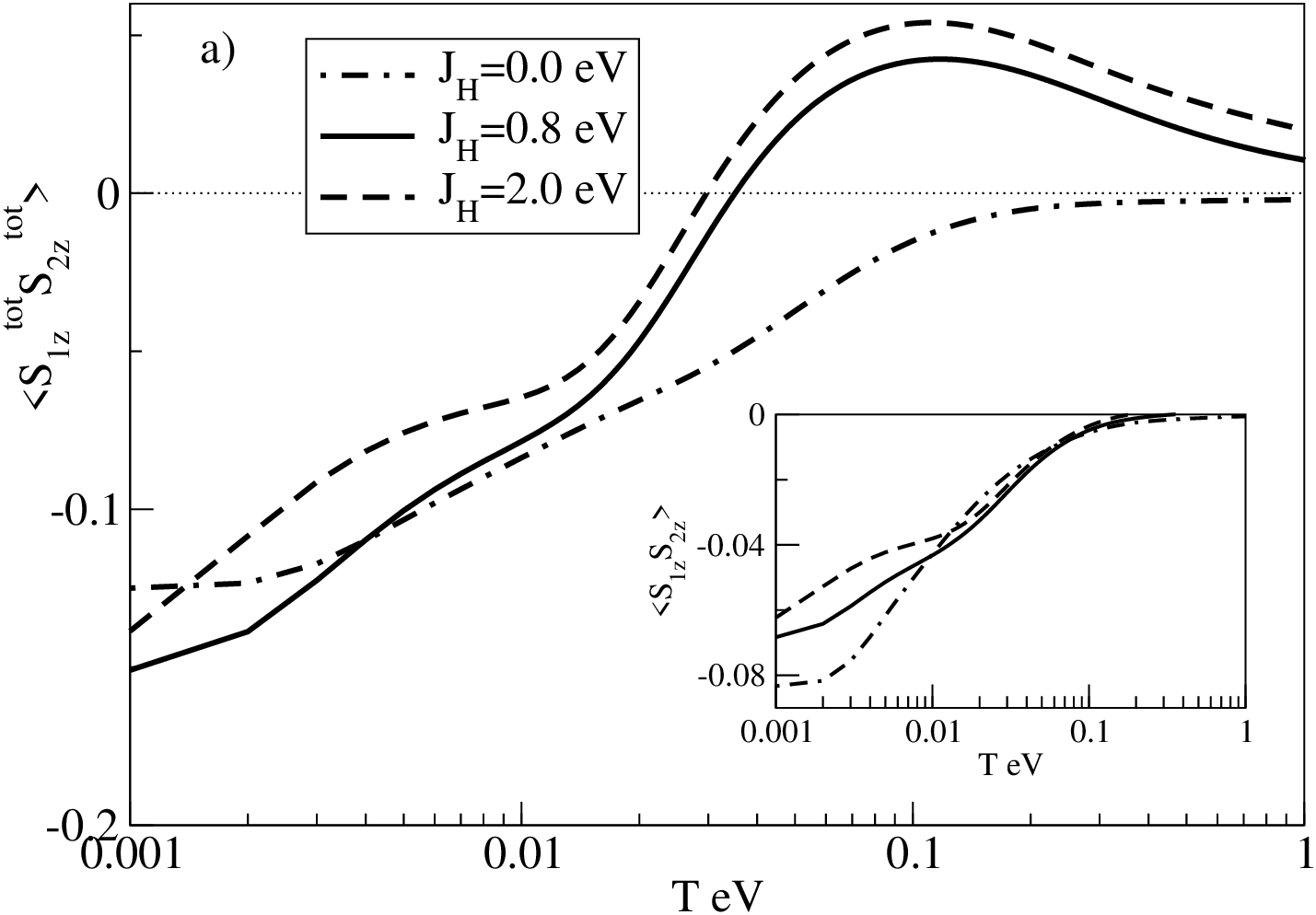}
\hspace{0.05\textwidth}
\includegraphics[angle=0, width=0.5\textwidth]{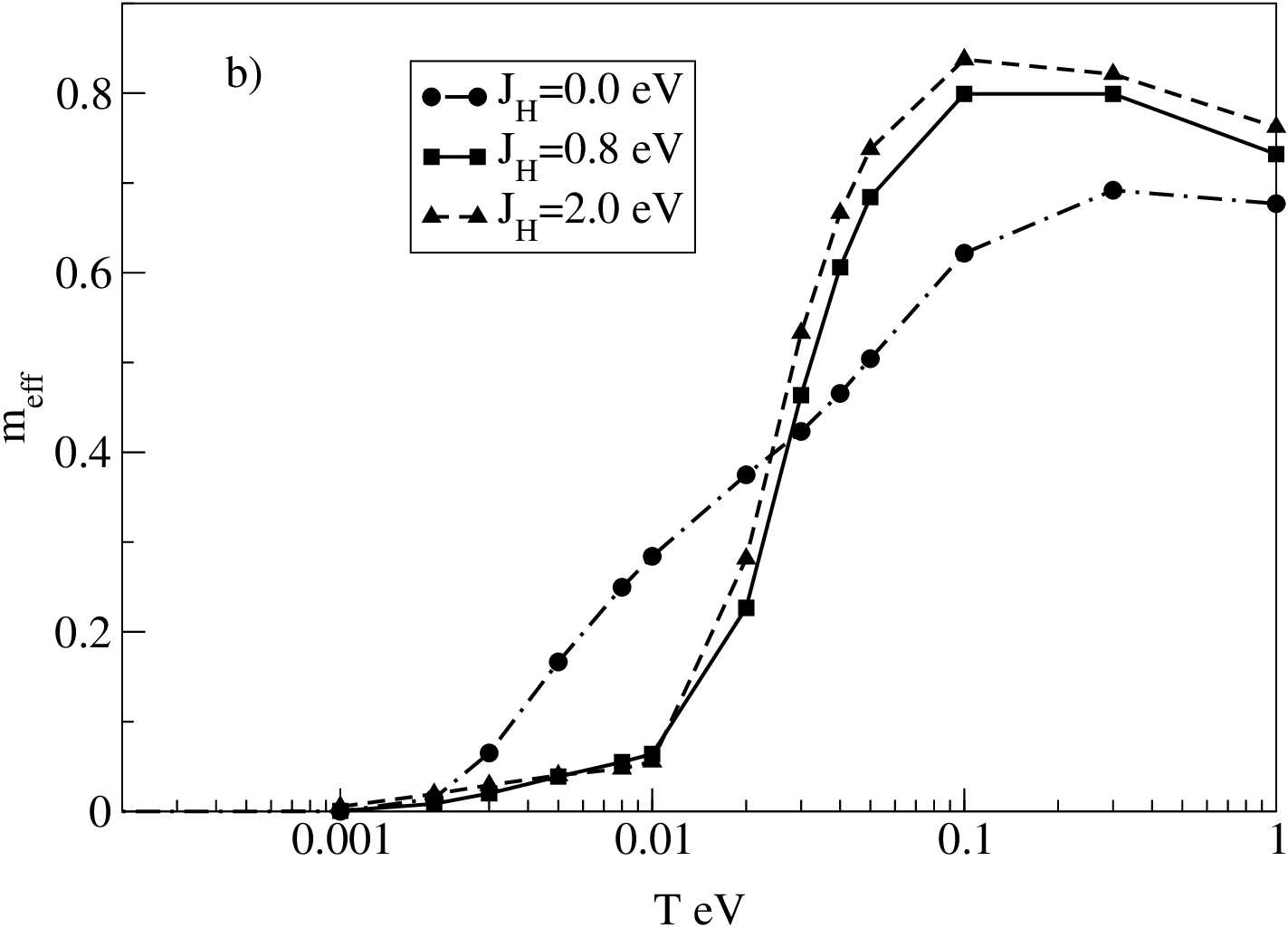}
\caption{\label{correlmagn} 
a) Temperature dependence of the total spin nearest neighbors correlation 
functions. The inset shows, for comparison, the correlations of only the 
localized spins. b) Temperature dependence of the local magnetic moment. 
On both a) and b), the temperature axis is logarithmic. }
\end{figure}

We discuss now the temperature dependence of the spin correlation functions 
at different values of Hund's coupling. We denote as 
${\bf S}^{tot}_i={\bf S}_i+{\sigma}_i$ the total spin at a given V ion. In 
Fig.~\ref{correlmagn}~a), we plot the nearest-neighbor spin correlation 
function $\langle {S}^{tot}_{1z}S^{tot}_{2z}\rangle$. At low temperatures the 
correlations are antiferromagnetic for the three values of $J_H$ considered. 
When the temperature is increased the correlations change from antiferro- to 
ferro-type at a temperature $T\sim 30$ meV for $J_H=0.8$ eV and $J_H=2.0$ eV, 
whereas it is always antiferromagnetic for $J_H=0.0$ eV. To emphasize the 
role of the itinerant electrons, we plot on the inset of 
Fig.~\ref{correlmagn}~a) the nearest-neighbor correlation functions 
$\langle {S}_{1z}S_{2z}\rangle$ of the localized spins only, which remain 
antiferromagnetic at any temperature and $J_H$.

This observation is in qualitative agreement with the inelastic neutron 
scattering measurements of Murani {\it et al.}~\cite{murani} where the 
correlations are antiferromagnetic (with a wave vector around $Q=0.6$ 
\AA$^{-1}$) below $T\approx 2$ K, and develop ferromagnetic-like 
correlations (corresponding to a peak at $Q=0$) with increasing temperature. 
A possible interpretation of these data could be the following: at low 
temperatures the ferromagnetic exchange, induced by the itinerant electrons 
due to the double exchange mechanism~\cite{zen}, is weaker than the direct 
antiferromagnetic exchange, and the resulting correlations are 
antiferromagnetic. With increasing temperature, carriers become more mobile, 
and, as a consequence, the effective ferromagnetic exchange grows. 
In Fig.~\ref{correlmagn}~b) we present the local effective moment per site, 
which is defined as: 
$m_{eff}(m_{eff}+1)\equiv\langle (\sum_i {\bf S}_{i}^{tot})^2\rangle/4$. 
At $J_H=0.0$ eV, when the localized spins are completely decoupled from the 
itinerant electrons, the magnetic moment is determined only by the localized 
spins. At zero temperature, the ground state is a singlet, and magnetic 
moment is zero. With increasing  temperature low lying triplet states become 
thermally populated, and this causes the formation  of the local moment. At 
high temperature the obtained value of 0.67 corresponds to a paramagnetic 
moment of 1.5 electrons in average on a V ion. \\
At $J_H=0.8$ eV and $J_H=2.0$ eV, the ground state has some triplet 
components but the weight of the singlet is still important. 
As a consequence, the effective moment is small at low temperature. 

Finally, we calculated the static magnetic susceptibility, defined as
$\chi (T)=\langle(S_{tot}^z)^2\rangle /T$. At high-temperature, for all 
values of the Hund's coupling the inverse susceptibilities  showing linear 
behavior , corresponding to a Curie-Weiss law. The Curie-Weiss temperatures 
extrapolated from high temperatures ($T\simeq 500 - 1000$ K ) are 
$\theta=-551$ K, -475 K and -360 K for $J_H=0$ eV, 0.8 eV and 2.0 eV, 
respectively. A detailed analysis of the magnetic susceptibility 
measurements~\cite{kondo} gives $\theta_{CW}\approx -37$ K when the fit is 
performed over the range $T\simeq 100-300$ K, but it gives $\theta\simeq -600$ 
K if fitted over the range $T\simeq 500 - 1000$ K. This latter value is of 
the same order as the one we have calculated, and we guess that the 
corresponding range of temperature is such that the correlations between 
tetrahedrons are negligible. According to our calculations, this temperature 
does mainly characterize the spin correlations but in 
the charge sectors $N=5$ and $N=7$. The value $\theta_{CW}\approx -37$ K, 
fitted at lower temperatures, would be more characteristic of spin 
correlations in the $N=6$ charge sector. 
\section{Conclusion}
We have performed an exact diagonalization of the small cluster, consisting 
of two disconnected tetrahedrons. The size of the cluster, constraint on the 
number of electrons and no exchange between two tetrahedrons gives us the 
possibility to make some qualitative description of the system, its spectrum  
and magnetic behavior. \\
We find that the ground state is a two fold degenerate singlet when the 
contribution from the itinerant electrons is not considered. The degeneracy 
is lifted and some important triplet components appear when the itinerant 
electrons are coupled to the localized spins by the Hund's exchange 
interaction with a realistic value. Consistent with experimental 
observations~\cite{murani}, the obtained temperature dependence of the spin 
correlation function shows a crossover from antiferromagnetic to ferromagnetic 
behavior when the temperature is increased. The increasing population of the 
magnetic states with temperature results in the formation of a non-zero 
residual magnetic moment which is also observed in the experiments. \\

We thank A.P. Murani, J. R. Iglesias and G. Jackeli for valuable discussions.



\begin{thebibliography}{20}
\bibitem{theory}
The theoretical study of the role of 
the geometrical frustration in $\rm{Li V_2 O_4}$ has been done 
by many authors (see, for example, Ref.~\cite{fulde,lacroix,laad}).
\end{thebibliography}
\end{document}